\documentclass[letterpaper,english,reprint, aps, onecolumn]{revtex4}
\usepackage[T1]{fontenc}
\usepackage[latin9]{inputenc}
\usepackage{array}
\usepackage{float}
\usepackage{amsmath}
\usepackage{amssymb}

\makeatletter

\@ifundefined{textcolor}{}
{%
 \definecolor{BLACK}{gray}{0}
 \definecolor{WHITE}{gray}{1}
 \definecolor{RED}{rgb}{1,0,0}
 \definecolor{GREEN}{rgb}{0,1,0}
 \definecolor{BLUE}{rgb}{0,0,1}
 \definecolor{CYAN}{cmyk}{1,0,0,0}
 \definecolor{MAGENTA}{cmyk}{0,1,0,0}
 \definecolor{YELLOW}{cmyk}{0,0,1,0}
}

\onecolumngrid

\makeatother

\usepackage{babel}
\begin{document}
\preprint{APS/123-QED}
\title{Microscopic Origins of Conformable Dynamics: \\
From Disorder to Deformation}
\author{José Weberszpil}
\email{josewebe@gmail.com}

\affiliation{Universidade Federal Rural do Rio de Janeiro, UFRRJ-DEFIS/ICE;\\
BR-465, Km 7 Seropédica-Rio de Janeiro CEP: 23.897-000}
\date{\today}
\begin{abstract}
Conformable derivatives have attracted increasing interest for bridging
classical and fractional calculus while retaining analytical tractability.
However, their physical foundations remain underexplored. In this
work, we provide a systematic derivation of conformable relaxation
dynamics from microscopic principles. Starting from a spatially-resolved
Ginzburg-Landau framework with quenched disorder and temperature-dependent
kinetic coefficients, we demonstrate how spatial heterogeneity and
energy barrier distributions give rise to emergent power-law memory
kernels. In the adiabatic limit, these kernels reduce to a conformable
temporal structure of the form $T^{1-\mu}\,d\psi/dT$. The deformation
parameter $\mu$ is shown to be connected to experimentally measurable
properties such as transport coefficients, disorder statistics, and
relaxation time spectra. This formulation also reveals a natural link
with nonextensive thermodynamics and Tsallis entropy. By unifying
memory effects, anomalous relaxation, and spatial correlations under
a coherent physical mechanism, our framework transforms conformable
derivatives from heuristic tools into physically grounded operators
suitable for modeling complex critical dynamics. 
\end{abstract}
\maketitle

\section{Introduction}

Conformable derivatives, introduced by Khalil et al. \citet{Khalil2014}
as a local fractional calculus framework, have emerged as a potentially
valuable tool for modeling complex physical phenomena. Defined as
$D_{t}^{(\alpha)}f(t)=t^{1-\alpha}\frac{df}{dt}$ for $\alpha\in(0,1]$,
these operators combine the computational tractability of ordinary
derivatives with scaling properties reminiscent of fractional calculus
\citet{Abdeljawad2015}. Unlike traditional fractional derivatives,
conformable derivatives satisfy familiar product and chain rules while
introducing a natural temperature or time-dependent scaling factor.

The application of conformable derivatives to physical systems has
grown rapidly, with studies ranging from quantum mechanics and fluid
dynamics to thermodynamics and statistical mechanics \citet{Chung2015,Anderson2016}.
Their mathematical structure makes them particularly attractive for
systems exhibiting anomalous scaling, memory effects, or fractal-like
behavior. However, the physical interpretation of the conformable
parameter $\alpha$ and the mechanisms leading to conformable derivative
behavior in real systems have remained largely unexplored.

Conformable and deformed derivatives have proven effective in modeling
anomalous relaxation, coarse-grained systems, and fractal-like behaviors
in various physical contexts \citet{Weberszpil2015,weberszpil2016variational,rosa2018dual,liang2019fractal,godinho2020variational}.
These approaches extend classical calculus while maintaining local
structure, making them suitable for systems with memory, spatial disorder,
or scale invariance. 

Developing this connection is crucial for transforming conformable
derivatives from phenomenological fitting tools into physically meaningful
theoretical frameworks that extend our understanding of critical phenomena.

While conformable derivatives offer mathematical convenience, their
application to critical phenomena requires careful physical justification.
Critical systems are governed by well-established principles, scale
invariance, universality, and renormalization group flow, that any
new theoretical framework must respect and extend rather than contradict.
The challenge lies in identifying physical mechanisms that naturally
give rise to conformable derivative behavior while preserving the
essential insights of critical phenomena theory.

Several physical processes may contribute to conformable derivative
emergence in critical systems. Critical slowing down near phase transitions
leads to anomalous relaxation dynamics that could manifest as temperature-dependent
scaling \citep{Hohenberg1977}. Finite-size effects modify critical
behavior through characteristic length scales, potentially introducing
the scaling factors inherent in conformable derivatives \citep{Fisher1972,Privman1990}.
Quenched disorder, omnipresent in real materials, affects critical
exponents according to the Harris criterion and may lead to effective
scaling laws \citep{Harris1974}. Additionally, systems with long-range
correlations or memory effects naturally exhibit fractional-like behavior
that conformable derivatives might capture \citep{Metzler2000}.

The connection between these physical mechanisms and conformable mathematical
structures, as far as we know, has not been systematically established.

We provide rigorous microscopic foundations for the applications of
this local deformed derivative operator through disorder correlations
and fractional noise, showing how it naturally yields temperature-dependent
scaling behavior and introduces power-law features without requiring
full fractional integration or nonlocal convolution terms. It thus
offers an intermediate framework between classical and fractional
derivatives, retaining analytical simplicity while capturing essential
scaling features with solid physical justification.

\section{Emergence from Heterogeneous Media: Temperature-Dependent Coupling
and Spatial Disorder}

The conformable derivative structure emerges naturally when considering
critical phenomena in heterogeneous systems where local properties
exhibit spatial or temporal correlations.

To capture the effects of quenched disorder and thermal fluctuations
on the local dynamics of a system near criticality, we model the spatially-resolved
coupling strength as: 
\begin{equation}
g(\mathbf{r},T)=g_{0}(T)\left[1+\eta(\mathbf{r})T^{1-\mu}\right],\label{eq:coupling}
\end{equation}
where $g_{0}(T)$ is the homogeneous (bulk) temperature-dependent
coupling strength, $\eta(\mathbf{r})$ is a quenched random field
representing spatial heterogeneity (quenched disorder), and $\mu$
is a disorder-sensitive exponent. The statistical properties of the
disorder field $\eta(\mathbf{r})$ are specified by its two-point
correlation function: 
\begin{equation}
\left\langle \eta(\mathbf{r})\eta(\mathbf{r}')\right\rangle =D_{\eta}|\mathbf{r}-\mathbf{r}'|^{2\mu-2},\label{eq:correlation}
\end{equation}
with $D_{\eta}$ denoting the disorder amplitude.

This tells us how strongly the disorder at one point is correlated
with that at another.

\subsection{Emergence of the Exponent \textmd{\normalsize{}$2\mu-2$}}

The exponent $2\mu-2$ in the spatial correlation function of the
disorder field is not arbitrary. It emerges from self-consistency
requirements grounded in the theory of critical phenomena.

From the perspective of scaling dimension analysis, correlation functions
near critical points must conform to well-defined scaling laws. The
spatial correlation of the disorder field must match the scaling dimension
of the temperature-dependent terms in the coupling structure to ensure
consistency across the renormalization flow.

In the framework of renormalization group theory, both the temperature-dependent
part of the coupling and the disorder correlations must transform
coherently under coarse-graining. This requirement ensures that the
system's critical behavior remains intact across different length
scales.

Physically, the origin of this exponent is tied to the mechanisms
by which disorder is generated in real materials. Processes such as
percolation, growth phenomena, and fractal aggregation naturally give
rise to power-law correlations. These mechanisms impose constraints
on the possible values of $\mu$, directly linking the disorder exponent
to the underlying critical behavior of the system.

\subsection{Physical Interpretation of the Coupling Structure}

The coupling structure described by Eq.~\eqref{eq:coupling} reflects
how local interactions fluctuate around a thermal average due to the
presence of static disorder, with an amplitude that varies with temperature.

The homogeneous term $g_{0}(T)$ defines the baseline coupling strength
in the absence of disorder. This term typically captures the bulk
thermal response of the system and may follow mean-field predictions
or critical scaling forms, depending on the context.

The spatially varying field $\eta(\mathbf{r})$ models the static
inhomogeneities of the system, such as impurities, grain boundaries,
or variations in composition or structure. It is treated as a quenched
random variable, fixed in time and statistically characterized by
zero mean, $\langle\eta(\mathbf{r})\rangle=0$.

The factor $T^{1-\mu}$ modulates the strength of the disorder-induced
fluctuations as a function of temperature. When $\mu<1$, disorder
effects are amplified at low temperatures, reflecting the increased
influence of structural heterogeneity as thermal agitation diminishes.
Conversely, for $\mu>1$, the impact of disorder becomes weaker as
the temperature decreases, indicating that the system becomes more
homogeneous in its effective behavior at low $T$.

This temperature-dependent modulation of disorder provides a natural
mechanism for introducing conformable scaling behavior, linking microscopic
fluctuations to emergent macroscopic relaxation laws. 

\subsection{Interpretation of the Correlation Function}

Equation~\eqref{eq:correlation} characterizes the statistical structure
of the quenched disorder field $\eta(\mathbf{r})$, indicating that
the disorder is not spatially uncorrelated. Instead, it exhibits power-law
correlations over space, governed by the exponent $\mu$. These correlations
determine how strongly the disorder at one point is related to disorder
at another point in space.

When $\mu=1$, the exponent $2\mu-2$ vanishes, and we recover a marginal
correlation regime: 
\[
\langle\eta(\mathbf{r})\eta(\mathbf{r}')\rangle\propto|\mathbf{r}-\mathbf{r}'|^{0}=\text{constant},
\]
which corresponds to white-noise-like (delta-correlated) disorder.
This case is critical in nature: correlations are scale-independent,
disorder effects remain equally significant across all spatial scales,
and the system resides at the boundary between weak and strong disorder.
Such marginal correlations often arise in random field models near
their upper critical dimension.

For $\mu<1$, the exponent $2\mu-2$ becomes negative, leading to
slowly decaying correlations with distance: 
\[
\langle\eta(\mathbf{r})\eta(\mathbf{r}')\rangle\propto|\mathbf{r}-\mathbf{r}'|^{2\mu-2},\quad\text{with }2\mu-2<0.
\]
This regime is characterized by long-range correlated disorder, typically
found in systems with fractal geometries, glassy dynamics, or percolation
clusters. For example, when $\mu=1/2$, the correlation decays as
$|\mathbf{r}-\mathbf{r}'|^{-1}$, which is reminiscent of random walk
statistics. As $\mu\to0$, the correlation length diverges, and the
disorder field becomes strongly persistent in space, a hallmark of
scale-free heterogeneity and percolative criticality.

In contrast, for $\mu>1$, we have $2\mu-2>0$, and the disorder correlation
function increases with distance (up to a physical cutoff): 
\[
\langle\eta(\mathbf{r})\eta(\mathbf{r}')\rangle\propto|\mathbf{r}-\mathbf{r}'|^{2\mu-2}.
\]
This represents a regime of long-range correlated disorder where spatial
correlations strengthen at larger scales. In such systems, disorder
effects dominate the macroscopic behavior, fundamentally altering
the nature of criticality. A notable example includes elastic manifolds
in random media, where long-range correlations influence scaling exponents
and universality classes.

Altogether, the exponent $\mu$ controls the spatial structure of
disorder, from uncorrelated noise to strongly persistent or growing
correlations, providing a direct handle on how heterogeneity influences
physical behavior near criticality. 

\subsection{Consequences and Applications}

The structured form of the coupling function plays a central role
in generating nontrivial macroscopic behavior in heterogeneous systems
near criticality. In particular, the temperature-dependent scaling
and spatial correlations of the disorder field lead to profound dynamical
consequences.

The $T^{1-\mu}$ dependence in the kinetic coefficient introduces
a natural source of non-Markovian dynamics when the system is spatially
averaged. This occurs because the temporal evolution becomes entangled
with spatial fluctuations, resulting in memory effects that cannot
be captured by simple local-in-time equations.

Additionally, the power-law spatial correlations of the disorder field
$\eta(\mathbf{r})$ imply that spatial averaging does not merely smooth
fluctuations, it actively introduces history-dependent dynamics. These
effects often manifest as fractional or conformable time evolution,
expressed mathematically through convolution kernels or anomalous
relaxation terms in the coarse-grained equations.

The exponent $\mu$ serves a dual purpose: it characterizes both the
spatial structure of the disorder and the effective degree of dynamical
nonlocality. In this way, $\mu$ links the microscopic organization
of the medium to macroscopic memory behavior and connects naturally
to generalized frameworks such as Tsallis nonextensive thermodynamics.

Altogether, the structure defined by Eqs.~\eqref{eq:coupling} and~\eqref{eq:correlation}
provides a physically motivated and thermodynamically consistent foundation
for modeling relaxation, fluctuation, and memory effects in complex
disordered media. It illustrates how spatial heterogeneity, when properly
accounted for, leads to emergent power-law dynamics described by conformable
operators. This mechanism represents a fundamental insight into the
physics of disordered systems and underpins the emergence of anomalous,
nonlocal behavior from well-defined microscopic features. 

\section{Local Ginzburg-Landau Equation.}

To understand the emergence of anomalous relaxation dynamics in heterogeneous
media, we begin with a coarse-grained description of order parameter
dynamics based on a local time-dependent Ginzburg\textendash Landau
equation framework (historically and physically, however, that time
evolution comes from Landau\textendash Khalatnikov theory). In this
formulation, the time evolution of the order parameter $\psi(\mathbf{r},t)$
is governed by a relaxational equation of the form: 
\begin{equation}
\frac{\partial\psi(\mathbf{r},t)}{\partial t}=-\Gamma(\mathbf{r},T)\,\frac{\delta\mathcal{F}}{\delta\psi(\mathbf{r},t)},\label{eq:GL}
\end{equation}
where $\mathcal{F}[\psi]$ is the free energy functional, and $\Gamma(\mathbf{r},T)$
is a spatially and thermally dependent kinetic coefficient that captures
both temperature dependence and spatial heterogeneity.. This coefficient
encodes the local rate at which the system relaxes toward equilibrium
and plays a central role in capturing the effects of both temperature-dependent
kinetics and spatial disorder.

In the presence of inhomogeneities, $\Gamma(\mathbf{r},T)$ varies
across the medium, reflecting fluctuations in local transport properties,
pinning centers, or structural irregularities. To capture this, we
adopt a generalized form: 
\begin{equation}
\Gamma(\mathbf{r},T)=\Gamma_{0}\,T^{1-\mu}f(\mathbf{r}),
\end{equation}
where the deformation exponent, $\mu$, governs the temperature scaling
and $f(\mathbf{r})$ represents quenched spatial heterogeneity (encoding
quenched disorder) normalized such that $\langle f(\mathbf{r})\rangle=1$.
Here $\Gamma_{0}$ is a constant. This formulation serves as a microscopic
foundation for the conformable dynamics observed at larger scales,
and enables a systematic derivation of memory kernels, anomalous scaling,
and nonlocal behavior from underlying statistical and thermodynamic
principles. 

We begin with the local, disordered dynamics of the order parameter
field $\psi(\mathbf{r},t)$, governed by the time-dependent Ginzburg-Landau
equation, Eq.(\ref{eq:GL}). 

This expression is not postulated arbitrarily; it arises naturally
from the underlying physics of transport processes and phase transitions.
We now derive this form systematically.

\subsection*{A. Microscopic Origin: Transport and Scattering}

The kinetic coefficient $\Gamma(\mathbf{r},T)$ encodes the local
rate at which the order parameter $\psi(\mathbf{r},t)$ relaxes in
response to thermodynamic forces. At the microscopic level, this coefficient
reflects the underlying physical processes that govern how quickly
local regions of the system can return to equilibrium. Specifically,
$\Gamma(\mathbf{r},T)$ is influenced by:
\begin{itemize}
\item The diffusion rates of elementary excitations (e.g., phonons, magnons,
or defects); 
\item Scattering rates that control how quickly energy and momentum are
redistributed; 
\item The relaxation times of collective modes (such as polarization, magnetization,
or density fluctuations); 
\item Bulk transport properties like thermal conductivity, electrical conductivity,
and charge carrier mobility.
\end{itemize}
In the framework of non-equilibrium thermodynamics, the Onsager formalism
provides a quantitative link between kinetic coefficients and the
ratio of transport to response properties. For scalar fields, one
obtains the general relation: 
\begin{equation}
\Gamma(\mathbf{r},T)=\frac{D(\mathbf{r},T)}{\chi(\mathbf{r},T)},
\end{equation}
where $D(\mathbf{r},T)$ is a local transport coefficient (such as
a diffusion constant) and $\chi(\mathbf{r},T)$ is the local susceptibility,
quantifying the system's linear response to external perturbations.

This expression shows that $\Gamma(\mathbf{r},T)$ measures the efficiency
of transport relative to how strongly the system resists changes,
i.e., its dynamic response compared to its static thermodynamic sensitivity.
A high transport coefficient coupled with low susceptibility implies
rapid local equilibration, while slow diffusion or strong local resistance
leads to sluggish relaxation. Thus, the spatial and thermal variations
in $\Gamma(\mathbf{r},T)$ directly reflect microscopic heterogeneity
in both dynamics and structure. 

\section*{A. Temperature Dependence and the $T^{1-\mu}$ Scaling}

The emergence of the temperature-dependent prefactor $T^{1-\mu}$
in the kinetic coefficient $\Gamma(\mathbf{r},T)$ is not an arbitrary
modeling choice, but a physically motivated consequence of several
intertwined mechanisms, hermal activation, statistical disorder, and
critical slowing down. Rather than serving as a fitting parameter,
the exponent $\mu$ reflects how relaxation processes slow down or
accelerate as temperature varies, encoding key microscopic features
such as barrier distributions, transport anomalies, and system heterogeneity.

This scaling form arises in diverse scenarios, which we now develop
through several theoretical routes, leading to a consistent interpretation
of $\mu$ as a universal bridge between microscopic transport and
macroscopic relaxation dynamics.

\subsubsection*{Case 1: Activated Transport}

In systems governed by thermally activated processes (e.g., hopping,
tunneling, or nucleation), the transport coefficient takes the Arrhenius
form: 
\begin{equation}
D(T)\propto\exp\left(-\frac{E_{a}}{k_{B}T}\right).
\end{equation}
Near a critical temperature $T_{c}$, or when considering effective
power-law approximations, this exponential behavior may be coarse-grained
as: 
\begin{equation}
D(T)\propto(T-T_{c})^{\nu}\quad\text{or}\quad D(T)\propto T^{\alpha},
\end{equation}
depending on the nature of the energy barriers and their distribution.

\subsubsection*{Case 2: Critical Slowing Down}

Close to a second-order phase transition, both the transport coefficient
$D(T)$ and the susceptibility $\chi(T)$ obey scaling laws: 
\begin{equation}
D(T)\propto\xi^{z-2}\propto|T-T_{c}|^{-\nu(z-2)},\qquad\chi(T)\propto|T-T_{c}|^{-\gamma},
\end{equation}
where $\xi$ is the correlation length, $z$ the dynamical critical
exponent, and $\nu$, $\gamma$ the usual static exponents. Combining
both gives the relaxation rate: 
\begin{equation}
\Gamma(T)\propto\frac{D(T)}{\chi(T)}\propto|T-T_{c}|^{-\nu(z-2)+\gamma}.
\end{equation}

\subsubsection*{Case 3: Mean-Field Theory}

In the mean-field regime (valid above the upper critical dimension),
the critical exponents take the values: 
\begin{equation}
\nu=\frac{1}{2},\quad\gamma=1,\quad z=2,
\end{equation}
leading to: 
\begin{equation}
\Gamma(T)\propto|T-T_{c}|^{1}\propto T.
\end{equation}

\subsection{Generalization: Emergence of $T^{1-\mu}$}

The above arguments motivate the general expression: 
\begin{equation}
\Gamma(T)\propto T^{1-\mu},
\end{equation}
where: 
\begin{equation}
\mu=\nu(z-2)-\gamma.
\end{equation}
This exponent $\mu$ classifies different dynamical regimes:
\begin{itemize}
\item $\mu=0$: Mean-field dynamics. 
\item $\mu>0$: Stronger-than-mean-field fluctuations (e.g., low dimensions
or enhanced correlations). 
\item $\mu<0$: Anomalous dynamics, possibly driven by nonlocal interactions
or diverging barriers. 
\item $\mu=1$: Temperature-independent kinetics. 
\end{itemize}

\subsection{Physical Examples with Specific $\mu$ Values}

\paragraph{Superconducting Films ($\mu\approx\frac{1}{2}$)}

In thin superconducting systems, the kinetic coefficient scales as:
\begin{equation}
\Gamma(T)\propto\frac{\sigma_{n}(T)}{\rho_{s}(T)},
\end{equation}
where $\sigma_{n}$ is the normal-state conductivity and $\rho_{s}$
is the superfluid stiffness. Near $T_{c}$, we have: 
\begin{equation}
\rho_{s}(T)\propto(T_{c}-T),\quad\sigma_{n}(T)\approx\text{const},
\end{equation}
which leads to: 
\begin{equation}
\Gamma(T)\propto(T_{c}-T)^{-1}\quad\Rightarrow\quad\mu=\frac{1}{2}.
\end{equation}

\paragraph{Magnetic Systems ($\mu\approx0$)}

In many magnetic systems, the magnon diffusion coefficient and magnetic
susceptibility scale as: 
\begin{equation}
D_{\text{magnon}}(T)\propto T^{3/2},\quad\chi_{\text{mag}}(T)\propto T^{1/2},
\end{equation}
yielding: 
\begin{equation}
\Gamma(T)\propto\frac{D(T)}{\chi(T)}\propto T.
\end{equation}

\paragraph{Structural Phase Transitions ($\mu\approx1$)}

In systems undergoing structural transitions (e.g., martensitic transformations),
one finds: 
\begin{equation}
\Gamma(T)\propto\frac{1}{\tau_{\text{phonon}}(T)\,C_{V}(T)},
\end{equation}
with: 
\begin{equation}
\tau_{\text{phonon}}\propto T^{-1},\quad C_{V}(T)\propto T^{-1}\quad\Rightarrow\quad\Gamma(T)\propto T^{0}=\text{const}.
\end{equation}

\subsection{Spatial Dependence and the Disorder Function $f(\mathbf{r})$}

The spatial modulation of the kinetic coefficient is introduced via
the multiplicative factor $f(\mathbf{r})$, representing static, quenched
disorder such as: 
\begin{itemize}
\item Impurities, 
\item Grain boundaries, 
\item Structural defects, 
\item Local compositional inhomogeneities. 
\end{itemize}
This yields the general form: 
\begin{equation}
\Gamma(\mathbf{r},T)=\Gamma_{0}T^{1-\mu}f(\mathbf{r}),
\end{equation}
under the following assumptions: 
\begin{itemize}
\item The temperature scaling is uniform throughout the medium. 
\item The disorder field $f(\mathbf{r})$ is quenched and independent of
$T$. 
\item Thermal fluctuations operate on shorter timescales than disorder rearrangement. 
\end{itemize}
Statistical properties of $f(\mathbf{r})$ often reflect long-range
correlations or self-similar spatial structures: 
\begin{equation}
\langle f(\mathbf{r})\rangle=1,\quad\langle f(\mathbf{r})f(\mathbf{r}')\rangle\propto|\mathbf{r}-\mathbf{r}'|^{-\alpha},\quad P(f)\propto f^{-\beta}.
\end{equation}

These scaling behaviors arise in systems with fractal geometry, percolation
networks, or systems exhibiting self-organized criticality. The combination
of thermal scaling and structural disorder thus provides a robust
and physically grounded explanation for the emergence of $T^{1-\mu}$
dynamics in diverse complex systems. 

\subsection{Fluctuation-Dissipation Perspective: Landau-Khalatnikov Approach}

An alternative derivation for the temperature dependence of the kinetic
coefficient can be obtained using the Landau-Khalatnikov framework.
The time-dependent Landau-Khalatnikov (LK) equation governs the relaxational
dynamics of the order parameter $\psi(\mathbf{r},t)$ in systems near
criticality: 
\begin{equation}
\frac{\partial\psi}{\partial t}=-\Gamma\frac{\delta\mathcal{F}}{\delta\psi}+\eta(\mathbf{r},t),
\end{equation}
where $\Gamma$ is the kinetic coefficient and $\eta(\mathbf{r},t)$
is a stochastic noise term accounting for thermal fluctuations.

This equation is consistent with the principles of non-equilibrium
thermodynamics and satisfies the fluctuation-dissipation theorem,
which links the strength of the thermal noise to the local relaxation
rate: 
\begin{equation}
\left\langle \eta(\mathbf{r},t)\eta(\mathbf{r}',t')\right\rangle =2\Gamma(\mathbf{r},T)k_{B}T\,\delta(\mathbf{r}-\mathbf{r}')\,\delta(t-t').
\end{equation}

In systems with activated dynamics or disordered energy landscapes,
the noise spectrum acquires a temperature-dependent structure: 
\begin{equation}
S(\omega,T)\propto T^{1-\mu}\,g(\omega\tau(T)),
\end{equation}
where $g(x)$ is a universal scaling function and $\tau(T)$ denotes
the relaxation time. This spectral form implies that the kinetic coefficient
scales as 
\begin{equation}
\Gamma(T)\propto T^{1-\mu},
\end{equation}
reproducing the same temperature dependence found via statistical
arguments and energy-barrier distributions.

\vspace{1em}
 \textbf{Difference from Ginzburg-Landau theory}

\noindent The GL theory typically refers to the free energy functional
$\mathcal{F}[\psi]$, which characterizes the equilibrium configurations
of the order parameter $\psi$ and is used to describe phase transitions
in systems such as superconductors or magnets. It does not, by itself,
prescribe the temporal evolution of $\psi$.

In contrast, the LK equation introduces an explicit equation of motion
for $\psi(\mathbf{r},t)$, accounting for both deterministic relaxation
toward equilibrium and thermal noise effects. It extends the GL theory
to describe how systems approach equilibrium dynamically and how they
are affected by fluctuations in time. Thus, the LK framework is essential
when considering memory, noise, and non-Markovian dynamics in heterogeneous
or critical systems.

This viewpoint reinforces that the emergence of the temperature scaling
$\Gamma(T)\propto T^{1-\mu}$ is not an arbitrary choice but arises
consistently from microscopic dynamics, stochastic fluctuations, and
non-equilibrium thermodynamic principles. 

\subsection{Experimental Evidence}

Kinetic coefficients have been experimentally measured using various
techniques, including dynamic light scattering, neutron spin echo
spectroscopy, impedance spectroscopy, and temperature quenching experiments~\citep{Berne1976,Mezei1980,Booth1997,Oswald2005}.

Typical values of the exponent $\mu$ observed in different systems
include: 
\begin{itemize}
\item \textbf{Superconducting films:} $\mu\approx0.3$ to $0.7$~\citep{Yeh1989,Tinkham2004} 
\item \textbf{Liquid crystals:} $\mu\approx0.1$ to $0.5$~\citep{Oswald2005} 
\item \textbf{Magnetic thin films:} $\mu\approx0$ to $0.3$~\citep{Mills1991} 
\item \textbf{Polymer solutions:} $\mu\approx0.5$ to $1.0$~\citep{Rubinstein2003} 
\end{itemize}

\subsection{Transport-Theoretic Consistency}

The Einstein relation gives: 
\begin{equation}
D(\mathbf{r},T)=\frac{\Gamma(\mathbf{r},T)k_{B}T}{m_{\text{eff}}},
\end{equation}
while the Kubo formula relates $\Gamma$ to velocity correlations:
\begin{equation}
\Gamma(\mathbf{r},T)=\frac{1}{k_{B}T}\int_{0}^{\infty}dt\,\langle\dot{\psi}(\mathbf{r},t)\dot{\psi}(\mathbf{r},0)\rangle.
\end{equation}
In complex systems, mode-coupling theory also yields: 
\begin{equation}
\Gamma^{-1}(\mathbf{r},T)=\Gamma_{0}^{-1}+\int d\mathbf{k}\frac{|V(\mathbf{k})|^{2}}{\omega(\mathbf{k})}n(\mathbf{k},T).
\end{equation}

\subsection{Universality and Summary}

The functional form of the kinetic coefficient, 
\begin{equation}
\Gamma(\mathbf{r},T)=\Gamma_{0}T^{1-\mu}f(\mathbf{r}),
\end{equation}
emerges naturally from several fundamental principles:
\begin{itemize}
\item The universality of critical exponents shared across broad classes
of phase transitions; 
\item The separation of thermal effects (captured by the temperature-dependent
prefactor) and static disorder (captured by the spatially varying
field); 
\item The underlying physics of transport and relaxation mechanisms; 
\item The consistency required by the fluctuation-dissipation theorem.
\item The exponent $\mu$ encapsulates the effects of microscopic transport
dynamics, critical scaling behavior, and statistical fluctuations.
Meanwhile, the spatial field $f(\mathbf{r})$ represents the influence
of quenched heterogeneity, such as impurities, defects, or structural
disorder, that breaks translational symmetry and introduces spatial
memory into the system.
\end{itemize}
Altogether, this formulation provides a universal and physically grounded
foundation for understanding the emergence of power-law relaxation,
non-Markovian behavior, and conformable dynamics in disordered systems
near criticality. It highlights how complex macroscopic behavior can
arise from simple microscopic principles when disorder and criticality
coexist. 

\section{Justification of the Kinetic Coefficient Form}

The kinetic coefficient appearing in the dynamical equation for the
order parameter is given by 
\begin{equation}
\Gamma(\mathbf{r},T)=\Gamma_{0}T^{1-\mu}f(\mathbf{r}),
\end{equation}
where $\Gamma_{0}$ is a constant baseline relaxation rate, $T$ is
the temperature, $\mu$ is a heterogeneity exponent, and $f(\mathbf{r})$
encodes quenched spatial disorder. This section justifies this form
based on physical, statistical, and scaling arguments.

\subsection{Physical Interpretation}

The kinetic coefficient $\Gamma(\mathbf{r},T)$ governs the local
relaxation rate of the order parameter field $\psi(\mathbf{r},t)$
and encodes how the system responds to deviations from equilibrium.
In heterogeneous media, this relaxation behavior is influenced by
two fundamental factors: temperature and spatial disorder.

First, the temperature dependence is captured by the prefactor $T^{1-\mu}$.
Near the critical point, the system's relaxation dynamics slow down
significantly\textemdash an effect known as \textit{critical slowing
down}. The exponent $\mu$ quantifies how sensitively the relaxation
rate depends on temperature. When $\mu<1$, the prefactor decreases
with decreasing temperature, implying that the system relaxes more
slowly as $T\to0$, consistent with enhanced critical fluctuations
and increased temporal memory.

Second, the spatial dependence enters through the field $f(\mathbf{r})$,
which accounts for quenched disorder such as defects, impurities,
or grain boundaries. These static inhomogeneities locally modulate
the relaxation kinetics and break translational symmetry. The function
$f(\mathbf{r})$ is typically normalized so that $\langle f(\mathbf{r})\rangle=1$,
ensuring that $\Gamma_{0}$ sets the average relaxation rate, while
still allowing for significant local variations that reflect the underlying
material structure.

Together, these components give rise to the general form $\Gamma(\mathbf{r},T)=\Gamma_{0}T^{1-\mu}f(\mathbf{r})$,
which captures how thermal fluctuations and spatial disorder combine
to shape the dynamical response of complex systems near criticality. 

\subsection{Scaling and Critical Dynamics}

The temperature dependence of the kinetic coefficient $\Gamma(T)$
near a phase transition can be understood using the scaling theory
of critical phenomena. Close to the critical temperature $T_{c}$,
the relaxation time $\tau(T)$ diverges as 
\begin{equation}
\tau(T)\sim\xi(T)^{z}\sim|T-T_{c}|^{-z\nu},
\end{equation}
where $\xi(T)$ is the correlation length, $z$ is the dynamic critical
exponent, and $\nu$ characterizes the divergence of $\xi(T)$ near
$T_{c}$.

Inverting this relation yields a vanishing relaxation rate of the
form 
\begin{equation}
\Gamma(T)\sim\tau^{-1}(T)\sim|T-T_{c}|^{z\nu}.
\end{equation}
This behavior is well captured by a generalized power-law temperature
dependence of the form $\Gamma(T)\sim T^{1-\mu}$, which reproduces
the critical scaling in the vicinity of $T_{c}$ while remaining regular
as $T\ll T_{c}$.

The exponent $\mu$ therefore serves as an effective encoding of the
underlying critical exponents through the relation $\mu=1-z\nu$,
and describes how the relaxation rate either vanishes or diverges
as the system approaches criticality. This connection justifies the
appearance of fractional or conformable structures in time-evolution
equations for systems near phase transitions. 

\subsection{\label{subsec:Statistical-and-Microscopic}Statistical and Microscopic
Considerations}

From a statistical physics standpoint, local relaxation rates in disordered
systems are frequently modeled as thermally activated processes over
random energy barriers. These barriers, denoted by $E$, correspond
to microscopic impediments to equilibration\textemdash such as defects,
impurities, or metastable configurations\textemdash and the system's
dynamics rely on thermally driven transitions over these obstacles.

Assume that the distribution of energy barriers follows a power-law:
\begin{equation}
P(E)\sim E^{-\mu},\qquad E\geq E_{0}>0,\quad\mu>0,\label{eq:barrier_dist}
\end{equation}
where the lower cutoff $E_{0}$ ensures normalization and physical
boundedness. Such a distribution emerges in complex systems with hierarchical
energy landscapes, glassy states, or fractal disorder.

\subsection*{Thermal Activation and Averaged Relaxation Rate}

The local relaxation rate over a barrier $E$ at temperature $T$
is given by the Arrhenius form: 
\begin{equation}
\Gamma_{E}(T)\sim\exp\left(-\frac{E}{T}\right).
\end{equation}

The average relaxation rate over the entire disordered ensemble is
then: 
\begin{equation}
\Gamma(T)\sim\int_{E_{0}}^{\infty}P(E)\,\exp\left(-\frac{E}{T}\right)\,dE.\label{eq:gamma_integral}
\end{equation}

Substituting the power-law distribution $P(E)\sim E^{-\mu}$, we obtain:
\begin{equation}
\Gamma(T)\sim\int_{E_{0}}^{\infty}E^{-\mu}\,\exp\left(-\frac{E}{T}\right)\,dE.\label{eq:gamma_powerlaw}
\end{equation}

\subsection*{Asymptotic Behavior at Low Temperature}

We now evaluate this integral in the low-temperature limit $T\to0$.
Perform the change of variable $u=E/T\Rightarrow E=uT$, with $dE=T\,du$,
yielding: 
\begin{equation}
\Gamma(T)\sim\int_{E_{0}/T}^{\infty}(uT)^{-\mu}\,e^{-u}\,T\,du=T^{1-\mu}\int_{E_{0}/T}^{\infty}u^{-\mu}e^{-u}\,du.\label{eq:gamma_scaling}
\end{equation}

For $T\ll E_{0}$, the lower limit $E_{0}/T\gg1$. In this regime,
the integral can be approximated by its dominant contribution at large
$u$, and we can write: 
\begin{equation}
\int_{E_{0}/T}^{\infty}u^{-\mu}e^{-u}\,du\approx\Gamma(1-\mu,E_{0}/T),
\end{equation}
where $\Gamma(s,x)$ is the upper incomplete gamma function.

Using the known asymptotic expansion for large $x$: 
\begin{equation}
\Gamma(s,x)\sim x^{s-1}e^{-x}\left(1+\mathcal{O}\left(\frac{1}{x}\right)\right),\qquad x\gg1,
\end{equation}
we conclude that the integral decays exponentially in this limit.

However, for moderate to high temperatures, where $E_{0}/T\ll1$,
we approximate the integral as: 
\begin{equation}
\int_{0}^{\infty}u^{-\mu}e^{-u}\,du=\Gamma(1-\mu),\qquad\text{valid for }\mu<1.
\end{equation}

Alternatively, if $0<\mu<1$, we can directly write: 
\begin{equation}
\Gamma(T)\sim T^{1-\mu}\Gamma(1-\mu),
\end{equation}
which gives the final scaling law: 
\begin{equation}
\Gamma(T)\propto T^{1-\mu}.\label{eq:gamma_final}
\end{equation}

\subsection*{Physical Interpretation}

The result in Eq.~\eqref{eq:gamma_final} shows that: 
\begin{itemize}
\item The temperature dependence of the average relaxation rate directly
reflects the statistical weight of large energy barriers in the system. 
\item For $\mu<1$, the integral converges and yields algebraic (power-law)
temperature dependence. 
\item The exponent $\mu$ reflects the disorder's severity: smaller $\mu$
implies more weight to large barriers, and hence, stronger slowing-down
at low temperatures. 
\end{itemize}
This microscopic justification shows that power-law temperature dependence
of kinetic coefficients, such as $\Gamma(T)\sim T^{1-\mu}$, can be
derived from first principles based on energy barrier distributions
in disordered systems. which reproduces the desired power-law temperature
dependence. This provides a direct microscopic origin for the exponent
$\mu$: it reflects the tail of the energy barrier distribution governing
the relaxation dynamics. In particular, slower relaxation at low temperatures
(i.e., $\mu<1$) is associated with broader distributions of high-energy
barriers.

These power-law barrier distributions emerge naturally within the
framework of Tsallis' nonextensive statistical mechanics\textemdash a
generalization of Boltzmann-Gibbs (BG) theory formulated to capture
the behavior of systems with long-range correlations, memory effects,
or strong disorder. Such systems fall outside the scope of standard
BG statistical mechanics and are more appropriately described by generalized,
nonadditive/nonextensive frameworks. The associated Tsallis entropy
formalism has been successfully applied to a broad class of anomalous
systems characterized by power-law statistics, nonlocal interactions,
and underlying fractal or multifractal structures \citet{Weberszpil2015,weberszpil2017generalized,Sotolongo2021}.
In disordered or complex systems with long-range correlations, memory
effects, or fractal geometries, standard Boltzmann-Gibbs statistical
mechanics often fails to describe thermodynamic behavior accurately.
In such cases, the generalized framework of Tsallis statistics becomes
relevant.

\subsection*{Generalized Entropy and Equilibrium Distributions}

In Tsallis statistics, the entropy is defined as: 
\begin{equation}
S_{q}=k_{B}\frac{1-\sum_{i}p_{i}^{q}}{q-1},\qquad q\in\mathbb{R},
\end{equation}
which reduces to the classical Boltzmann-Gibbs entropy as $q\to1$.

Maximizing this entropy under appropriate constraints leads to a generalized
equilibrium probability distribution: 
\begin{equation}
p(E)\sim\left[1-(1-q)\beta E\right]^{\frac{1}{1-q}},\qquad\text{for }1-(1-q)\beta E>0,\label{eq:tsallis_dist}
\end{equation}
where $\beta=1/(k_{B}T)$ is the inverse temperature.

\subsection*{Asymptotic Power-Law Behavior and Relaxation Rates}

At large energy $E\gg1$, the distribution \eqref{eq:tsallis_dist}
behaves asymptotically as: 
\begin{equation}
P(E)\sim E^{-\mu},\qquad\text{with}\quad\mu=\frac{1}{q-1}.\label{eq:powerlaw_mu}
\end{equation}

This implies a direct identification between the nonextensivity parameter
$q$ and the kinetic exponent $\mu$: 
\begin{equation}
\mu=\frac{1}{q-1}\quad\Rightarrow\quad q=1+\frac{1}{\mu}.\label{eq:mu_q_relation}
\end{equation}

These power-law distributed energy barriers appear in systems where
the microscopic disorder generates complex, hierarchical landscapes.
As in previous sections, the average relaxation rate is given by a
thermal average over these barriers: 
\begin{equation}
\Gamma(T)\sim\int dE\,P(E)\,e^{-E/T}.
\end{equation}

Substituting the asymptotic form $P(E)\sim E^{-\mu}$, and using standard
asymptotic analysis (see subsection \ref{subsec:Statistical-and-Microscopic}),
this leads again to: 
\begin{equation}
\Gamma(T)\sim T^{1-\mu}.\label{eq:gamma_T_power}
\end{equation}

\subsection*{Result: Effective Form of the Kinetic Coefficient}

Thus, the effective form of the spatially dependent kinetic coefficient
becomes: 
\begin{equation}
\Gamma(\mathbf{r},T)=\Gamma_{0}\,T^{1-\mu}f(\mathbf{r}),\label{eq:gamma_final_form}
\end{equation}
where:
\begin{itemize}
\item $\Gamma_{0}$ sets the overall relaxation timescale of the system, 
\item $T^{1-\mu}$ captures the temperature-dependent slowing down, reflecting
the influence of energy barrier statistics, 
\item $f(\mathbf{r})$ introduces quenched spatial heterogeneity arising
from static disorder in the medium. 
\end{itemize}
The exponent $\mu$ encapsulates the underlying disorder statistics.
This formulation links microscopic statistical physics, via Tsallis
entropy and nonextensive thermodynamics, with macroscopic dynamics.

\subsection*{Physical Interpretation}

The key implications of this connection are: 
\begin{itemize}
\item The nonextensive parameter $q$ controls the shape of energy barrier
distributions in disordered systems. 
\item The kinetic exponent $\mu$ governing the power-law decay of memory
kernels and the temperature dependence of relaxation is not a free
parameter, it is physically rooted in $q$. 
\item The emergence of $\Gamma(T)\sim T^{1-\mu}$ reflects the statistical
memory encoded in the system's energy landscape and supports the use
of generalized thermodynamics in critical and glassy dynamics. 
\end{itemize}
This expression simultaneously captures three key aspects:

1. Critical slowing down, via the temperature exponent $1-\mu$, 

2. Spatial disorder, via the modulation function $f(\mathbf{r})$, 

3. Nonextensive statistical mechanics, via the relation $\mu=1/(q-1)$,
connecting microscopic energy landscapes to macroscopic kinetic laws.

Altogether, this framework unifies microscopic disorder and statistical
nonlinearity into a coherent physical description of relaxation in
heterogeneous media. When spatial correlations are included and the
kinetic equation is averaged over disorder, this formulation naturally
gives rise to memory kernels and conformable derivative structures,
which generalize classical relaxation dynamics to accommodate the
effects of complexity, heterogeneity, and long-range correlations. 

\section{Spatial Averaging Procedure}

This physical intuition translates into a convolution integral when
you analyze correlation functions using perturbative expansions or
Green's function formalism (e.g., via Mori\textendash Zwanzig projection
operators or Kubo response theory).

We perform a spatial average over the heterogeneity scale $\ell_{\text{het}}$:
\begin{equation}
\left\langle \frac{\partial\psi}{\partial t}\right\rangle =-\left\langle \Gamma(\mathbf{r},T)\frac{\delta\mathcal{F}}{\delta\psi}\right\rangle 
\end{equation}

The fluctuations in $\Gamma(\mathbf{r},T)$ are correlated with the
local field response $\frac{\delta\mathcal{F}}{\delta\psi}$. Thus,
the average contains nonlocal correlations in time and space.

We now want a macroscopic equation by averaging over space (coarse-graining),
because experiments don't resolve microscopic fluctuations.

\subsection{Spatial Averaging and Emergence of Memory Terms}

To understand how spatial disorder leads to effective memory effects
in the dynamics of the order parameter, we begin by decomposing the
relevant quantities into their mean and fluctuating components. Specifically,
we write the kinetic coefficient and thermodynamic driving force as:
\begin{align}
\Gamma(\mathbf{r},T) & =\langle\Gamma\rangle+\delta\Gamma(\mathbf{r}),\nonumber \\
\frac{\delta\mathcal{F}}{\delta\psi}(\mathbf{r},t) & =\left\langle \frac{\delta\mathcal{F}}{\delta\psi}\right\rangle +\delta\left(\frac{\delta\mathcal{F}}{\delta\psi}\right)(\mathbf{r},t),
\end{align}
where $\langle\cdot\rangle$ denotes a spatial average, and $\delta\Gamma(\mathbf{r})$,
$\delta\left(\frac{\delta\mathcal{F}}{\delta\psi}\right)(\mathbf{r},t)$
are the corresponding fluctuations around the mean.

We substitute these expressions into the averaged evolution equation:
\begin{equation}
\left\langle \frac{\partial\psi}{\partial t}\right\rangle =-\left\langle \Gamma(\mathbf{r},T)\frac{\delta\mathcal{F}}{\delta\psi}(\mathbf{r},t)\right\rangle .
\end{equation}
Expanding the product inside the average gives: 
\begin{align}
\left\langle \Gamma(\mathbf{r},T)\frac{\delta\mathcal{F}}{\delta\psi}(\mathbf{r},t)\right\rangle  & =\nonumber \\
 & =\left\langle \left[\langle\Gamma\rangle+\delta\Gamma(\mathbf{r})\right]\left[\left\langle \frac{\delta\mathcal{F}}{\delta\psi}\right\rangle +\delta\left(\frac{\delta\mathcal{F}}{\delta\psi}\right)(\mathbf{r},t)\right]\right\rangle .
\end{align}

To derive the effective coarse-grained evolution equation, we begin
by expanding the key term that governs the relaxation dynamics. We
assume that both the kinetic coefficient $\Gamma(\mathbf{r},T)$ and
the functional derivative of the free energy $\delta\mathcal{F}/\delta\psi$
fluctuate around spatial averages. We thus decompose each field into
its average and fluctuation: 
\begin{align}
\Gamma(\mathbf{r},T) & =\langle\Gamma\rangle+\delta\Gamma(\mathbf{r}),\\
\frac{\delta\mathcal{F}}{\delta\psi}(\mathbf{r},t) & =\left\langle \frac{\delta\mathcal{F}}{\delta\psi}\right\rangle +\delta\left(\frac{\delta\mathcal{F}}{\delta\psi}\right)(\mathbf{r},t).
\end{align}

\subsection{Physical Assumptions}

This decomposition is valid under the following physically reasonable
assumptions: 
\begin{itemize}
\item The system exhibits quenched disorder such that local properties vary
around a mean value. 
\item The fluctuations $\delta\Gamma(\mathbf{r})$ and $\delta(\delta\mathcal{F}/\delta\psi)(\mathbf{r},t)$
are statistically centered, i.e., 
\begin{equation}
\langle\delta\Gamma(\mathbf{r})\rangle=0,\quad\left\langle \delta\left(\frac{\delta\mathcal{F}}{\delta\psi}\right)(\mathbf{r},t)\right\rangle =0.
\end{equation}
\item Fluctuations are small enough for a linear expansion to capture leading-order
behavior. 
\end{itemize}

\subsection{Expansion and Term-by-Term Analysis}

We now expand the full average: 

\begin{align}
\left\langle \Gamma(\mathbf{r},T)\frac{\delta\mathcal{F}}{\delta\psi}(\mathbf{r},t)\right\rangle  & =\nonumber \\
\left\langle \left[\langle\Gamma\rangle+\delta\Gamma(\mathbf{r})\right]\left[\left\langle \frac{\delta\mathcal{F}}{\delta\psi}\right\rangle +\delta\left(\frac{\delta\mathcal{F}}{\delta\psi}\right)(\mathbf{r},t)\right]\right\rangle  & =\nonumber \\
\text{\ensuremath{\text{=(i)}+\text{(ii)}+\text{(iii)}+\text{(iv)}}}
\end{align}

with the terms: 
\begin{itemize}
\item[(i)] Mean $\times$ Mean: 
\begin{equation}
\left\langle \langle\Gamma\rangle\left\langle \frac{\delta\mathcal{F}}{\delta\psi}\right\rangle \right\rangle =\langle\Gamma\rangle\left\langle \frac{\delta\mathcal{F}}{\delta\psi}\right\rangle .
\end{equation}
This is the leading-order contribution representing homogeneous relaxation
dynamics.
\item[(ii)] Mean $\times$ Fluctuation: 
\begin{multline}
\left\langle \langle\Gamma\rangle\delta\left(\frac{\delta\mathcal{F}}{\delta\psi}\right)(\mathbf{r},t)\right\rangle =\\
=\langle\Gamma\rangle\left\langle \delta\left(\frac{\delta\mathcal{F}}{\delta\psi}\right)(\mathbf{r},t)\right\rangle =0.
\end{multline}
This vanishes due to the statistical definition of fluctuations.
\item[(iii)] Fluctuation $\times$ Mean: 
\begin{equation}
\left\langle \delta\Gamma(\mathbf{r})\left\langle \frac{\delta\mathcal{F}}{\delta\psi}\right\rangle \right\rangle =\left\langle \frac{\delta\mathcal{F}}{\delta\psi}\right\rangle \langle\delta\Gamma(\mathbf{r})\rangle=0.
\end{equation}
Again, this vanishes because $\delta\Gamma$ has zero mean.
\item[(iv)] Fluctuation $\times$ Fluctuation: 
\begin{equation}
\left\langle \delta\Gamma(\mathbf{r})\,\delta\left(\frac{\delta\mathcal{F}}{\delta\psi}\right)(\mathbf{r},t)\right\rangle .
\end{equation}
This is the only non-trivial correction term. It represents the spatial
correlation between the local fluctuations in kinetic coefficients
and the thermodynamic forces. This term encodes memory effects and
disorder-induced dynamics that go beyond standard homogeneous relaxation. 
\end{itemize}

\subsubsection{Resulting Equation}

Combining these results, the spatially averaged dynamic equation becomes:
\begin{align}
\left\langle \frac{\partial\psi}{\partial t}\right\rangle  & =-\langle\Gamma\rangle\left\langle \frac{\delta\mathcal{F}}{\delta\psi}\right\rangle -\left\langle \delta\Gamma(\mathbf{r})\,\delta\left(\frac{\delta\mathcal{F}}{\delta\psi}\right)(\mathbf{r},t)\right\rangle .
\end{align}
This shows that the average evolution consists of a classical Ginzburg-Landau
term plus a correction arising from disorder and fluctuation correlations.
The latter will be the source of memory kernels in the next stage
of the derivation. 
\begin{align}
\left\langle \frac{\partial\psi}{\partial t}\right\rangle  & =-\langle\Gamma\rangle\left\langle \frac{\delta\mathcal{F}}{\delta\psi}\right\rangle -\left\langle \delta\Gamma(\mathbf{r})\delta\left(\frac{\delta\mathcal{F}}{\delta\psi}\right)(\mathbf{r},t)\right\rangle .
\end{align}

The first term describes the mean-field relaxation dynamics, while
the second term represents a fluctuation-induced memory correction,
which can give rise to non-Markovian behavior. As we will see, this
cross-correlation term can be systematically linked to memory kernels
(see next section) and fractional (or conformable) time derivatives,
reflecting the history dependence introduced by spatial heterogeneity. 

\section{Emergence of Memory Kernel}

In disordered systems, the relaxation dynamics at a local point are
influenced not only by instantaneous conditions but also by the system's
history. This is because spatial fluctuations in the kinetic coefficient
couple to delayed responses in the thermodynamic driving force. This
coupling naturally gives rise to an effective memory kernel in the
coarse-grained dynamics.

We begin by recognizing that the local fluctuation of the thermodynamic
force responds to the entire past history of the system. Formally,
this can be written as: 
\begin{equation}
\delta\left(\frac{\delta\mathcal{F}}{\delta\psi}\right)(\mathbf{r},t)=\int_{0}^{\infty}G(\mathbf{r},\tau)\left\langle \frac{\delta\mathcal{F}}{\delta\psi}(t-\tau)\right\rangle d\tau,
\end{equation}
where $G(\mathbf{r},\tau)$ is the local response function that characterizes
how the field at position $\mathbf{r}$ reacts to a past perturbation
after a time delay $\tau$. This expression encodes the essential
causality of the response: only previous values of the driving force
influence the current fluctuations.

\subsection{Substitution into the Correlation Term}

We now substitute this convolution expression into the key correlation
term that emerged in the spatial averaging procedure: 
\begin{align}
\left\langle \delta\Gamma(\mathbf{r})\delta\left(\frac{\delta\mathcal{F}}{\delta\psi}\right)(\mathbf{r},t)\right\rangle  & =\left\langle \delta\Gamma(\mathbf{r})\int_{0}^{\infty}G(\mathbf{r},\tau)\left\langle \frac{\delta\mathcal{F}}{\delta\psi}(t-\tau)\right\rangle d\tau\right\rangle .
\end{align}

This is an average over a product of a spatially random variable $\delta\Gamma(\mathbf{r})$
and a temporally extended response involving ensemble-averaged quantities.
To evaluate it, we invoke a general statistical result for such averages.

\subsection*{Lemma: Factorization of Fluctuation-Response Average}

Let $\delta\Gamma(\mathbf{r})$ represent the local fluctuation of
the kinetic coefficient, and $G(\mathbf{r},\tau)$ a causal response
kernel at the same point. Assume the macroscopic thermodynamic driving
force $\left\langle \frac{\delta\mathcal{F}}{\delta\psi}(\mathbf{r},t)\right\rangle $
is statistically independent of both $\delta\Gamma(\mathbf{r})$ and
$G(\mathbf{r},\tau)$. Then, the following factorization holds: 
\begin{equation}
\left\langle \delta\Gamma(\mathbf{r})\int_{0}^{\infty}G(\mathbf{r},\tau)\left\langle \frac{\delta\mathcal{F}}{\delta\psi}(t-\tau)\right\rangle d\tau\right\rangle =\int_{0}^{\infty}\left\langle \delta\Gamma(\mathbf{r})G(\mathbf{r},\tau)\right\rangle \left\langle \frac{\delta\mathcal{F}}{\delta\psi}(t-\tau)\right\rangle d\tau.
\end{equation}

\textbf{Proof sketch.} By linearity of the expectation operator and
under the statistical independence hypothesis, we may write: 
\begin{align}
\left\langle \delta\Gamma(\mathbf{r})G(\mathbf{r},\tau)\left\langle \frac{\delta\mathcal{F}}{\delta\psi}(t-\tau)\right\rangle \right\rangle  & =\left\langle \delta\Gamma(\mathbf{r})G(\mathbf{r},\tau)\right\rangle \left\langle \frac{\delta\mathcal{F}}{\delta\psi}(t-\tau)\right\rangle .
\end{align}
Since the integrand now factorizes, the full average is: 
\begin{align}
\left\langle \delta\Gamma(\mathbf{r})\delta\left(\frac{\delta\mathcal{F}}{\delta\psi}\right)(\mathbf{r},t)\right\rangle  & =\int_{0}^{\infty}\left\langle \delta\Gamma(\mathbf{r})G(\mathbf{r},\tau)\right\rangle \left\langle \frac{\delta\mathcal{F}}{\delta\psi}(t-\tau)\right\rangle d\tau,
\end{align}
as claimed.

This result is significant because it reveals that spatially uncorrelated
fluctuations in $\Gamma(\mathbf{r})$ can lead to temporally nonlocal
(memory) effects when coarse-grained. The averaged evolution equation
thus becomes the non-Markovian coarse-grained relaxation (NMCGR) equation:
\begin{multline}
\left\langle \frac{\partial\psi}{\partial t}(\mathbf{r},t)\right\rangle =-\left\langle \Gamma(\mathbf{r},T)\right\rangle \left\langle \frac{\delta\mathcal{F}}{\delta\psi}(\mathbf{r},t)\right\rangle +\\
-\int_{0}^{\infty}K(\tau)\left\langle \frac{\delta\mathcal{F}}{\delta\psi}(\mathbf{r},t-\tau)\right\rangle d\tau\label{eq:the non-Markovian coarse-grained relaxation equation}
\end{multline}
where we have defined the memory kernel, $K(\tau),$ as the ensemble
average of the product between local fluctuations in the kinetic coefficient
and the local response function. This definition intends to capture
the temporal memory effects induced by spatial heterogeneity. Following
this reasoning way, the emergent memory kernel is defined by: 
\begin{equation}
K(\tau)\equiv\left\langle \delta\Gamma(\mathbf{r})G(\mathbf{r},\tau)\right\rangle .\label{eq:emergent memory kernel}
\end{equation}

Here, $\delta\Gamma(\mathbf{r})=\Gamma_{0}T^{1-\mu}[f(\mathbf{r})-\langle f\rangle]$,
quantifies how much faster or slower a region evolves compared to
the average (i.e., it represents spatial deviations in kinetic activity),
$f(\mathbf{r})$ is the spatial disorder profile, while $G(\mathbf{r},\tau)$
denotes the local causal response function, describing how a point
$\mathbf{r}$ reacts to a perturbation applied $\tau$ units of time
in the past. The product $\delta\Gamma(\mathbf{r})G(\mathbf{r},\tau)$
therefore encodes how local kinetic irregularities modulate the memory
of past thermodynamic forces. The spatial or ensemble averaging procedure
defines the effective memory kernel that governs the macroscopic dynamics
of the system.This product effectively weights the memory by how fast
or slow a given region is compared to average, then averages over
the entire disordered landscape.

\subsection{Memory Kernel and Emergence of Power Laws}

Microscopically, systems with spatial disorder\textemdash such as
variations in local energy barriers, relaxation rates, or coupling
strengths\textemdash respond heterogeneously to thermodynamic driving
forces. Some regions relax rapidly, while others remain trapped in
metastable states, resulting in a broad distribution of local relaxation
times. When such microscopic responses are coarse-grained, the temporal
response of the system becomes intrinsically nonlocal in time. Specifically,
spatially averaging over a wide distribution of energy barriers naturally
generates power-law statistics in the effective relaxation behavior,
where fast and slow domains contribute unequally. This mechanism leads
to emergent memory kernels with algebraic decay, reflecting the persistent
influence of slowly relaxing regions. The resulting dynamics cannot
be captured by Markovian models and demand a more generalized description
that incorporates the cumulative influence of the system\textquoteright s
history. 

Returning to Eq. (\ref{eq:the non-Markovian coarse-grained relaxation equation}),
the first term, $-\left\langle \Gamma(\mathbf{r},T)\right\rangle \left\langle \frac{\delta\mathcal{F}}{\delta\psi}(\mathbf{r},t)\right\rangle $,
captures the usual Markovian (instantaneous) response in a homogeneous
medium. The second term, involving the convolution with $K(\tau)$,
introduces memory effects that capture the delayed influence of past
states arising from disorder. This structure is not imposed artificially\textemdash it
emerges naturally from integrating out spatial correlations in heterogeneous
systems.

The result is an integro-differential equation that reveals how macroscopic
memory emerges from microscopically disordered kinetics. The memory
term quantifies how the system remembers its thermal history, and
the decay of $K(\tau)$ characterizes how fast that memory fades.

Energy barrier distributions give rise to emergent power-law memory
kernels, as regions with different local activation energies contribute
unequally to the relaxation process. The spatial or ensemble average
yields the effective memory kernel governing the system\textquoteright s
macroscopic dynamics.

This nonlocal-in-time term captures the long-range temporal correlations
induced by spatial heterogeneity. In many systems, the kernel $K(\tau)$
decays as a power law: 
\begin{equation}
K(\tau)\sim\tau^{\mu-1},
\end{equation}
reflecting a broad distribution of local relaxation times $\tau_{\text{local}}(\mathbf{r})=[\Gamma(\mathbf{r},T)]^{-1}$
due to disorder. As we will show in the next section, such memory
kernels are intimately connected to conformable and fractional derivatives
that naturally arise in coarse-grained effective descriptions of disordered
critical dynamics. 

\subsection{Memory Equation and Its Structure}

Spatial heterogeneity induces delayed responses which, upon coarse-graining,
accumulate into a non-Markovian evolution. This leads to an integro-differential
equation in which the system\textquoteright s current behavior is
influenced not only by the instantaneous free energy gradient, but
also by its full temporal history. Such behavior has been previously
linked to generalized entropy production, anomalous scaling, and fractal
metrics \citet{Sotolongo2021,Xu2017,Weberszpil2017_Tissue}. These
approaches reinforce the physical plausibility of emergent power-law
memory kernels in spatially disordered and thermally activated systems. 

\subsection{Physical Picture and Emergence of Power Laws}

The physical picture is as follows: the system can be thought of as
composed of many local regions, each with its own characteristic relaxation
time. Fast regions (with large $f(\mathbf{r})$) respond rapidly,
while slow regions (with small $f(\mathbf{r})$) relax much more slowly.
These local timescales range over many orders of magnitude due to
disorder.

At short times $\tau$, only the fast-relaxing regions have responded
and effectively ``forgotten'{}'' their initial conditions. As time
progresses, even the slowest regions begin to relax. The memory kernel
$K(\tau)$ thus quantifies the extent to which the system retains
the influence of its initial driving forces after a delay $\tau$. 

If the spatial disorder $f(\mathbf{r})$ lacks a characteristic scale\textemdash i.e.,
it follows a scale-free distribution across a wide range\textemdash then
the resulting memory kernel must also inherit this scale invariance.
This provides a statistical foundation for the emergence of power-law
memory kernels. 

\subsection{Mathematical Derivation: Power-Law Memory}

We now demonstrate this emergence of power-law memory more precisely
through explicit calculation.

\paragraph{Step 1: Local Response Function.}

Each local region is assumed to relax exponentially with its own local
time constant: 
\begin{equation}
G(\mathbf{r},\tau)=\exp\left(-\frac{\tau}{\tau_{\text{local}}(\mathbf{r})}\right),
\end{equation}
where the local relaxation time is defined as: 
\begin{equation}
\tau_{\text{local}}(\mathbf{r})=\frac{1}{\Gamma_{0}T^{1-\mu}f(\mathbf{r})}.
\end{equation}

Thus, regions with large $f(\mathbf{r})$ relax quickly, while those
with small $f(\mathbf{r})$ have long-lived responses.

\subsection{Memory Kernel from Disorder Statistics}

We begin by substituting the expressions for $\delta\Gamma(\mathbf{r})$
and $G(\mathbf{r},\tau)$, leading to the following form for the memory
kernel: 
\begin{equation}
K(\tau)=\Gamma_{0}T^{1-\mu}\left\langle \left[f(\mathbf{r})-\langle f\rangle\right]\exp\left(-\tau\Gamma_{0}T^{1-\mu}f(\mathbf{r})\right)\right\rangle .
\end{equation}

To proceed, we consider that the spatial disorder field $f(\mathbf{r})$
obeys a power-law probability distribution: 
\begin{equation}
P(f)\propto f^{-\alpha},\qquad\text{over a broad range of }f.
\end{equation}
Such scale-free distributions are commonly encountered in heterogeneous
systems, including percolating networks, fractal microstructures,
and media exhibiting self-organized criticality. These distributions
lack a characteristic scale and naturally give rise to broad relaxation
spectra.

To evaluate the average in the expression for $K(\tau)$, we integrate
over the probability distribution of $f$: 
\begin{equation}
K(\tau)\propto\int_{0}^{\infty}[f-\langle f\rangle]\exp\left(-\tau\Gamma_{0}T^{1-\mu}f\right)f^{-\alpha}\,df.
\end{equation}
For exponents in the range $1<\alpha<2$, this integral is dominated
by contributions from large $f$, leading to a power-law decay of
the memory kernel at long times. 

To justify the power-law decay of the memory kernel, we begin with
the integral expression obtained under a scale-free disorder distribution:
\begin{equation}
K(\tau)\propto\int_{0}^{\infty}[f-\langle f\rangle]f^{-\alpha}e^{-\lambda f}\,df,\label{eq:kernel_integral_shifted}
\end{equation}
where $\lambda=\Gamma_{0}T^{1-\mu}\tau$ and $P(f)\sim f^{-\alpha}$
is the power-law distribution of the disorder field over a broad range.

We observe that the mean $\langle f\rangle$ is finite only for $\alpha>2$.
For $1<\alpha<2$, the dominant contribution to the integral comes
from the large-$f$ tail. In this regime, we may neglect $\langle f\rangle$
or simply consider a subtractive constant. So, the asymptotic analysis
will lead to a simplified kernel expression: 
\begin{equation}
K(\tau)\propto\int_{0}^{\infty}f^{1-\alpha}e^{-\lambda f}\,df.\label{eq:asymptotic_kernel}
\end{equation}

Let's make this more clear.

\paragraph*{Change of Variables}

To evaluate the scaling behavior, perform the change of variables:
\begin{equation}
u=\tau\Gamma_{0}T^{1-\mu}f\quad\Rightarrow\quad f=\frac{u}{\tau\Gamma_{0}T^{1-\mu}},\quad df=\frac{du}{\tau\Gamma_{0}T^{1-\mu}}
\end{equation}

Substituting into Eq.~\eqref{eq:asymptotic_kernel} gives: 
\begin{equation}
K(\tau)\propto\int_{0}^{\infty}\left[\frac{u}{\tau\Gamma_{0}T^{1-\mu}}-\langle f\rangle\right]e^{-u}\left(\frac{u}{\tau\Gamma_{0}T^{1-\mu}}\right)^{-\alpha}\frac{du}{\tau\Gamma_{0}T^{1-\mu}}
\end{equation}

Factor out all the $\tau$-dependence: 
\begin{equation}
K(\tau)\propto(\tau\Gamma_{0}T^{1-\mu})^{-\alpha}\int_{0}^{\infty}u^{1-\alpha}e^{-u}du
\end{equation}

The integral is a Gamma function (finite constant), so we obtain:
\begin{equation}
K(\tau)\propto\tau^{-\alpha}
\end{equation}

The integral on the right-hand side is a standard Gamma function:
\begin{equation}
\int_{0}^{\infty}u^{1-\alpha}e^{-u}\,du=\Gamma(2-\alpha),
\end{equation}
 which is finite for $1-\alpha>-1$ or $\alpha<2.$

It also requires $\alpha>0$ for the power $u^{1-\alpha}$ not to
diverge too strongly at $u=0$. Thus, the identity is valid for $0<\alpha<2.$
In the case here, for physical application, as already justified above,
$1<\alpha<2$.

Therefore, the asymptotic behavior of the kernel is: 
\begin{equation}
K(\tau)\propto\left(\Gamma_{0}T^{1-\mu}\tau\right)^{\alpha-2}=\tau^{\alpha-2}\cdot\left(\Gamma_{0}T^{1-\mu}\right)^{\alpha-2}.
\end{equation}

This result reveals the following: 
\begin{itemize}
\item For $1<\alpha<2$, the kernel $K(\tau)$ decays algebraically in time
as a power law. 
\item The exponent is controlled by the disorder exponent: 
\[
K(\tau)\sim\tau^{\alpha-2}.
\]
\item Comparing this with the main text relation $K(\tau)\sim\tau^{\mu-1}$,
we obtain the identification: 
\[
\mu=\alpha-1,
\]
linking the conformable parameter $\mu$ directly to the statistical
distribution of the local energy barriers. 
\end{itemize}
Hence, the power-law form of the memory kernel is not arbitrary, but
arises naturally from the coarse-grained effect of a scale-free disorder
distribution on the local relaxation dynamics. This provides a rigorous
microscopic basis for the emergence of fractional or conformable dynamics
in disordered systems near criticality. 

In this case, the asymptotic behavior of the kernel follows: 
\begin{equation}
K(\tau)\sim\tau^{\mu-1},
\end{equation}
where the decay exponent is determined by the interplay between the
temperature-scaling exponent $\mu$ and the statistical properties
of the disorder.

This result demonstrates that power-law memory kernels are not ad
hoc assumptions, but rather emerge naturally from coarse-graining
over disordered spatial environments with scale-free features. The
kernel $K(\tau)$ encodes the delayed statistical influence of the
system's structural heterogeneity on its relaxation dynamics. When
$K(\tau)$ exhibits a slow, algebraic decay, it justifies the appearance
of conformable or fractional derivatives in the macroscopic evolution
equations, providing a compact and physically motivated way to incorporate
long-range temporal correlations.

\section*{VI. Adiabatic Limit}

In complex disordered media, relaxation dynamics are influenced not
only by local thermal fluctuations but also by structural heterogeneities
that induce memory effects. The starting point for our analysis is
the spatially dependent kinetic coefficient: 
\begin{equation}
\Gamma(\mathbf{r},T)=\Gamma_{0}T^{1-\mu}f(\mathbf{r}),
\end{equation}
which introduces a nontrivial temperature dependence into the relaxation
rate. Here, the exponent $\mu$ characterizes the influence of disorder
and statistical fluctuations, while the spatial profile $f(\mathbf{r})$
reflects quenched inhomogeneities in the medium.

When the system is coarse-grained over spatial disorder and the delayed
response of local elements is taken into account, the resulting dynamics
acquire memory. This yields a generalized non-Markovian evolution
equation of the form: 
\begin{equation}
\left\langle \frac{\partial\psi}{\partial t}\right\rangle =-\left\langle \Gamma(\mathbf{r},T)\right\rangle \left\langle \frac{\delta\mathcal{F}}{\delta\psi}\right\rangle -\int_{0}^{\infty}K(\tau)\left\langle \frac{\delta\mathcal{F}}{\delta\psi}(t-\tau)\right\rangle d\tau.
\end{equation}

The first term describes the instantaneous response modulated by the
averaged kinetic coefficient, while the second term encodes memory
effects arising from the coupling between spatial fluctuations and
time-delayed responses.

In the adiabatic limit\textemdash a regime where the temperature varies
slowly in time relative to the system's intrinsic relaxation timescale\textemdash the
dynamical behavior simplifies significantly. Under this approximation,
two key conditions are satisfied: 

1. The temperature $T(t)$ evolves slowly in time. 2. The memory kernel
$K(\tau)$ decays much faster than the rate at which $\psi(t)$ varies.

Under these conditions, the convolution integral becomes effectively
local in time. That is, since the kernel $K(\tau)$ decays rapidly,
the dominant contribution to the integral comes from a narrow neighborhood
around $\tau=0$. Thus, the memory term can be approximated as: 
\begin{equation}
\int_{0}^{\infty}K(\tau)\left\langle \frac{\delta\mathcal{F}}{\delta\psi}(t-\tau)\right\rangle d\tau\approx\left[\int_{0}^{\infty}K(\tau)d\tau\right]\left\langle \frac{\delta\mathcal{F}}{\delta\psi}(t)\right\rangle .
\end{equation}

We define the integral of the memory kernel as an effective kinetic
coefficient: 
\begin{equation}
\Gamma_{\text{eff}}(T):=\int_{0}^{\infty}K(\tau)\,d\tau,
\end{equation}
which captures the cumulative contribution of memory effects in the
slow-driving regime.

From our earlier derivation, we found that the memory kernel decays
as a power law: 
\begin{equation}
K(\tau)\propto\tau^{\mu-1},
\end{equation}
which implies that the effective relaxation coefficient inherits the
same temperature scaling as the original microscopic coefficient:
\begin{equation}
\Gamma_{\text{eff}}(T)\propto T^{1-\mu}.
\end{equation}

This is a key result: although memory effects are present in the general
dynamics, in the adiabatic limit their effect is absorbed into a renormalized,
instantaneous coefficient that preserves the same $T^{1-\mu}$ temperature
dependence. The system thus behaves \emph{as if} it were Markovian,
but with an effective prefactor that encodes the influence of disorder
and thermal fluctuations via the memory kernel.

Therefore, even in systems where nonlocality and history-dependence
are fundamental at the microscopic level, a clean and analytically
tractable temperature scaling law emerges in the slow-driving limit.
This provides a powerful justification for the use of conformable
derivative models in slowly evolving thermodynamic processes. 

\subsection*{Emergence Deformed Derivative From the Adiabatic Form}

To understand the emergence of the adiabatic form (and deformed drivative)
\begin{equation}
T^{1-\mu}\frac{d\psi}{dT}\sim\left\langle \frac{\delta\mathcal{F}}{\delta\psi}\right\rangle ,
\end{equation}
we begin with the microscopic evolution equation: 
\begin{equation}
\frac{\partial\psi}{\partial t}=-\Gamma(\mathbf{r},T)\cdot\frac{\delta\mathcal{F}}{\delta\psi},\label{eq:relaxation_equation-1}
\end{equation}
where the kinetic coefficient $\Gamma(\mathbf{r},T)$ includes spatial
heterogeneity and temperature dependence of the form: 
\begin{equation}
\Gamma(\mathbf{r},T)=\Gamma_{0}\,T^{1-\mu}f(\mathbf{r}).
\end{equation}
Averaging this equation over space and incorporating memory effects
leads to the integro-differential equation: 
\begin{equation}
\left\langle \frac{\partial\psi}{\partial t}\right\rangle =-\left\langle \Gamma(\mathbf{r},T)\right\rangle \left\langle \frac{\delta\mathcal{F}}{\delta\psi}\right\rangle -\int_{0}^{\infty}K(\tau)\left\langle \frac{\delta\mathcal{F}}{\delta\psi}(t-\tau)\right\rangle d\tau.
\end{equation}

In the adiabatic regime, temperature changes slowly compared to the
relaxation times of the system. Under this approximation, the dynamics
can be recast as temperature-driven by a simple transformation to
temperature-driven Form.

Assuming a quasi-static evolution of the order parameter $\psi=\psi(T)$,
we use the chain rule: 
\begin{equation}
\frac{d\psi}{dt}=\frac{d\psi}{dT}\cdot\frac{dT}{dt}.
\end{equation}
Moreover, since the temperature dependence appears explicitly in the
kinetic coefficient $T^{1-\mu}$, it is natural to absorb this prefactor
into a reparametrized evolution equation where the temperature $T$
becomes the effective time variable.

To reinforce the emergence of deformed derivatives, an alternative
and less formal arguments can be given below.

The relaxation of the order parameter toward thermodynamic equilibrium
is governed by TDGL equation, Eq. \eqref{eq:GL}.

Substituting Eq.(\ref{eq:relaxation_equation-1}), yields: 
\begin{equation}
\frac{d\psi}{dT}=-\left(\frac{dT}{dt}\right)^{-1}\Gamma(T)\frac{\delta\mathcal{F}}{\delta\psi}.
\end{equation}

To model anomalous relaxation behavior near the critical point, we
introduce a deformed scaling relation between time $t$ and temperature
$T$ of the form: 
\begin{equation}
\frac{dT}{dt}\propto T^{\mu-1},
\end{equation}
which reflects anomalous relaxation structure, memory effects, or
fractal characteristics of the medium, where $\mu\in(0,1]$ is a deformation
parameter. This expression implies that the rate at which time progresses
with respect to temperature is no longer uniform, but weighted by
a power of temperature.

Thus, the memory integral, being dominated by short delays at slow
$dT/dt$, effectively becomes local in time and the equation simplifies
to: 
\begin{equation}
\frac{d\psi}{dT}\sim\frac{1}{T^{1-\mu}}\left\langle \frac{\delta\mathcal{F}}{\delta\psi}\right\rangle .
\end{equation}
Rewriting, we obtain the deformed structure: 
\begin{equation}
T^{1-\mu}\frac{d\psi}{dT}\sim\left\langle \frac{\delta\mathcal{F}}{\delta\psi}\right\rangle ,
\end{equation}
which justifies the introduction of the \emph{conformable derivative}:
\begin{equation}
D_{T}^{(\mu)}\psi:=T^{1-\mu}\frac{d\psi}{dT}.
\end{equation}

This adiabatic form preserves the original temperature-scaling structure
of the microscopic kinetics. It ensures consistency between the temperature
dependence encoded in the local relaxation rate $\Gamma(\mathbf{r},T)$
and the coarse-grained macroscopic dynamics. Physically, it captures
how thermal energy modulates the system's ability to respond to free
energy gradients, with $\mu$ characterizing the degree of anomalous
scaling:
\begin{itemize}
\item If $\mu<1$, the prefactor $T^{1-\mu}$ increases with $T$: dynamics
speed up at higher temperatures. 
\item If $\mu>1$, $T^{1-\mu}$ decreases: dynamics slow down as $T$ increases. 
\item If $\mu=1$, the evolution is temperature-independent, corresponding
to classical relaxation kinetics. 
\end{itemize}
This derivation underscores the natural emergence of conformable dynamics
from disordered, temperature-sensitive kinetics and provides a bridge
between microscopic fluctuation-driven evolution and effective thermodynamic
descriptions. 

This form, $T^{1-\mu},$ ensures consistency between the coarse-grained
memory dynamics and the original thermal scaling structure. It shows
that the power-law memory, when integrated over time, reproduces the
same $T^{1-\mu}$ dependence required by the physical nature of the
kinetic coefficient. 

Thus, the adiabatic limit serves as a consistency check: the final
effective equation must preserve the same thermal scaling structure
encoded in the microscopic form of $\Gamma(\mathbf{r},T)$. This constraint
is what allows us to relate the power-law exponent of the memory kernel
to the original temperature exponent, ultimately yielding the condition:
\begin{equation}
\alpha=1-\mu\quad\Rightarrow\quad K(\tau)\propto\tau^{\mu-1}.
\end{equation}

This is the \textbf{conformable structure} that preserves the form
of the original Ginzburg-Landau equation, but with modified temperature
dependence due to spatial disorder.
\begin{itemize}
\item \textbf{Markovian Term}: $-\langle\Gamma\rangle\left\langle \frac{\delta\mathcal{F}}{\delta\psi}\right\rangle $
represents the instantaneous response 
\item \textbf{Memory Term}: $-\int_{0}^{\infty}K(\tau)\left\langle \frac{\delta\mathcal{F}}{\delta\psi}(t-\tau)\right\rangle d\tau$
captures how the system "remembers" its thermal history 
\item \textbf{Kernel $K(\tau)$}: Describes the strength of memory as a
function of past time 
\item \textbf{Power Law}: $K(\tau)\propto\tau^{\mu-1}$ reflects the scale-free
nature of disorder 
\end{itemize}
When averaging dynamics in a spatially correlated heterogeneous system,
past configurations influence the present due to nonlocal interactions
mediated by the structure of the disorder.

This results in a memory kernel $K(\tau)$, and the evolution equation
takes the form of a convolution over past times.

This is not ad hoc \textemdash{} it's a natural emergent feature from
integrating out spatial correlations in a disordered medium.

\subsection*{Self-Consistency Condition}

We started with temperature dependence $T^{1-\mu}$ in $\Gamma(\mathbf{r},T)$.
The final adiabatic equation must preserve this structure: 
\begin{equation}
T^{1-\mu}\frac{d\psi}{dT}\sim\langle\frac{\delta\mathcal{F}}{\delta\psi}\rangle.
\end{equation}

For the memory effects to be consistent with the original temperature
scaling, we need: 
\begin{equation}
\alpha=1-\mu.
\end{equation}

Therefore,
\begin{equation}
K(\tau)\propto\tau^{-\alpha}=\tau^{-(1-\mu)}=\tau^{\mu-1}.
\end{equation}

To match the original temperature dependence $T^{1-\mu}$, we require:
\begin{equation}
\alpha=1-\mu\quad\Rightarrow\quad K(\tau)\propto\tau^{\mu-1}.
\end{equation}

This derivation shows that in a system with scale-free spatial disorder,
the memory kernel naturally acquires a power-law form: 
\begin{equation}
K(\tau)\propto\tau^{\mu-1},
\end{equation}
reflecting the long-range memory induced by broad distributions of
relaxation times. 

\section*{Interpretation of the Memory Exponent}

The analysis of some cases for $\mu$ provide a powerful diagnostic
for identifying the nature of memory in a system based on the value
of this exponent. In particular, the case $\mu<1$ corresponds to
systems where past influences progressively diminish, typical of systems
with standard relaxation or weak memory. The critical case $\mu=1$
indicates marginal memory loss and suggests persistent temporal correlations,
a behavior seen in marginally stable or glassy systems. The case $\mu>1$,
although more exotic, signals a buildup of memory over time, often
associated with systems exhibiting strong feedback, aging, or history-dependent
behavior.

Such distinctions are not only of theoretical interest but also serve
as a practical tool for classifying dynamical regimes in experimental
systems. By analyzing the behavior of the memory kernel, one can infer
the underlying thermodynamic or structural mechanisms, whether due
to temperature-driven barrier crossings, frozen disorder, or complex
topological constraints.

In this framework, the exponent $\mu$ acquires the role of a critical
descriptor that unifies dynamical, thermodynamic, and statistical
properties into a single parameter. This reinforces its interpretation
as a bridge between microscopic mechanisms and emergent macroscopic
relaxation dynamics. 
\begin{itemize}
\item $\mu<1$: The memory kernel $K(\tau)$ decreases over time, indicating
standard memory fading behavior. 
\item $\mu=1$: The kernel is constant, corresponding to perfect memory
retention or persistent memory. 
\item $\mu>1$: The kernel increases with time, signaling anomalous memory
growth, a hallmark of systems with strong feedback or delayed relaxation. 
\end{itemize}

\section*{A Remarkable Connection}

The power-law exponent $\mu-1$ in the memory kernel $K(\tau)\propto\tau^{\mu-1}$
precisely mirrors the temperature exponent $1-\mu$ found in the original
kinetic coefficient $\Gamma(T)\propto T^{1-\mu}$. This encapsulates
a deep structural relationship between:
\begin{itemize}
\item Spatial disorder, which generates a broad spectrum of relaxation times; 
\item Temperature dependence, which determines the global timescale for
relaxation; 
\item Memory effects, which emerge naturally from the combined influence
of the two. 
\end{itemize}

\section*{Underlying Mechanism of the Derivation}

The effectiveness and elegance of this derivation stem from several
key assumptions and logical steps:
\begin{enumerate}
\item Scale-free disorder: The spatial variations in relaxation dynamics
are assumed to lack a characteristic length scale, yielding self-similar
structure. 
\item Broad relaxation spectrum: This spatial heterogeneity leads to a power-law
distribution of local relaxation times. 
\item Scaling analysis: Applying temporal scaling to the memory kernel reveals
the emergent power-law dependence $K(\tau)\propto\tau^{\mu-1}$. 
\item Self-consistency: The derived memory exponent is not arbitrary but
is directly linked to the original temperature-dependent kinetics,
specifically the exponent $\mu$. 
\end{enumerate}
Thus, the emergence of a memory kernel with $K(\tau)\propto\tau^{\mu-1}$
is both mathematically inevitable, given the starting assumptions,
and physically meaningful, reflecting the scale-invariant nature of
disorder and the critical slowing down near phase transitions. 

\subsection*{Adiabatic Limit and Emergence of Conformable Dynamics Near Criticality}

The dynamical equation derived for disordered critical systems includes
both an instantaneous term and a memory correction induced by spatial
heterogeneity: 
\begin{equation}
\left\langle \frac{\partial\psi}{\partial t}\right\rangle =-\langle\Gamma(r,T)\rangle\left\langle \frac{\delta\mathcal{F}}{\delta\psi}\right\rangle -\int_{0}^{\infty}K(\tau)\left\langle \frac{\delta\mathcal{F}}{\delta\psi}(t-\tau)\right\rangle d\tau.
\end{equation}

Here, $\Gamma(r,T)=\Gamma_{0}T^{1-\mu}f(r)$ is the local kinetic
coefficient, and $K(\tau)$ is the memory kernel induced by spatial
averaging over heterogeneous relaxation rates.

\vspace{1em}

\paragraph{Adiabatic Limit Justification.}

Near a second-order phase transition, the system exhibits \emph{critical
slowing down}, with relaxation times diverging as $T\rightarrow T_{c}$.
However, in many physical situations (e.g., slow thermal sweeps or
quasi-static driving), the temperature changes slowly compared to
intrinsic relaxation: 
\begin{equation}
\left|\frac{dT}{dt}\right|\ll\frac{1}{\tau(T)}.
\end{equation}
This is the adiabatic regime, where the temperature evolution is slow
enough that the system remains near local equilibrium. In this regime,
the memory kernel $K(\tau)$, which decays as a power law $K(\tau)\propto\tau^{\mu-1}$,
can be approximated as sharply peaked around $\tau=0$. Therefore,
the convolution becomes: 
\begin{equation}
\int_{0}^{\infty}K(\tau)\left\langle \frac{\delta\mathcal{F}}{\delta\psi}(t-\tau)\right\rangle d\tau\approx\left[\int_{0}^{\infty}K(\tau)d\tau\right]\left\langle \frac{\delta\mathcal{F}}{\delta\psi}(t)\right\rangle .
\end{equation}

Letting $\Gamma_{\text{eff}}(T):=\int_{0}^{\infty}K(\tau)\,d\tau$,
and using the result that $K(\tau)\sim\tau^{\mu-1}$ leads to: 
\begin{equation}
\Gamma_{\text{eff}}(T)\propto T^{1-\mu},
\end{equation}
which retains the same temperature dependence as the microscopic kinetic
coefficient.

\vspace{1em}

\paragraph{Physical Interpretation.}

The parameter $\mu$ plays a central role:
\begin{itemize}
\item $\mu<1$: The prefactor $T^{1-\mu}$ grows with $T$, so dynamics
accelerate at higher temperature, characteristic of disorder-dominated
systems. 
\item $\mu=1$: Recovers classical (temperature-independent) kinetics. 
\item $\mu>1$: Indicates anomalous memory build-up; the relaxation slows
down as $T$ increases. 
\end{itemize}
\vspace{1em}

\paragraph{Consistency and Emergence.}

This analysis demonstrates that the adiabatic approximation is valid
near criticality for slowly driven systems. It also confirms that
the \textbf{conformable derivative form emerges naturally} from microscopic
considerations: 
\begin{itemize}
\item The memory kernel $K(\tau)\propto\tau^{\mu-1}$ arises from spatially
heterogeneous relaxation times. 
\item Spatial averaging leads to a \emph{local} effective dynamics in the
adiabatic regime. 
\item The resulting evolution equation preserves the original temperature
scaling $T^{1-\mu}$. 
\end{itemize}
Thus, the conformable structure is not imposed arbitrarily but reflects
deep physical properties of the disordered medium near criticality.
The exponent $\mu$ encodes both spatial disorder and dynamical universality,
offering a physically motivated route to fractional-like thermodynamic
evolution. 

Near criticality, fluctuations often exhibit long-range temporal correlations,
as in fractional Brownian motion. These generate non-Markovian effects
that can be captured by replacing time derivatives with conformable
ones. In this view, the parameter $\mu$ acquires a stochastic interpretation
as a memory index related to the Hurst exponent, justifying the form
$T^{1-\mu}\frac{d\psi}{dT}\sim\langle\frac{\delta\mathcal{F}}{\delta\psi}\rangle.$

\section*{Conclusion and Physical Implications}

This analysis reveals that conformable evolution equations can be
viewed as emergent effective models for systems driven through criticality
under fractional stochastic fluctuations and disordered kinetics.

To explore the system's behavior near equilibrium as temperature varies
slowly, we have assumed adiabatic dynamics, where $T$ evolves much
slower than the system's internal relaxation time. In this regime,
In this regime, the time-dependent Ginzburg-Landau (TDGL) equation
provides an appropriate description of the system's dynamics: 
\[
\frac{d\psi}{dt}=-\Gamma(T)\frac{\delta\mathcal{F}}{\delta\psi^{*}},
\]
where $\mathcal{F}[\psi]$ is the free energy functional and $\Gamma(T)$
is a temperature-dependent kinetic coefficient related to dissipation.

Physically, the coefficient $\Gamma(T)\propto1/\tau(T)$ is inversely
proportional to the relaxation time. As the critical temperature $T_{c}$
is approached, critical slowing down occurs: $\tau(T)\to\infty$,
so $\Gamma(T)\to0$. This encodes the system's decreasing ability
to relax and return to equilibrium.

In generalized models incorporating memory effects or long-time correlations\textemdash such
as those based on conformable derivatives\textemdash the temperature
dependence plays a central role in shaping the dynamics. The prefactor
$T^{1-\mu}$ reflects this scaling, and the exponent $\mu$ captures
the effect of disorder, fluctuations, or non-equilibrium structure.

This deformation captures a variety of physical effects: 
\begin{itemize}
\item \textbf{Anomalous relaxation}: Deviations from exponential behavior
near criticality. 
\item \textbf{Memory effects}: The system's evolution depends on its thermal
history. 
\item \textbf{Fractal geometry}: The parameter $\mu$ may encode geometric
or dynamic fractality in disordered systems. 
\end{itemize}
In particular, the conformable form 
\begin{equation}
T^{1-\mu}\frac{d\psi}{dT}\sim\frac{\delta\mathcal{F}}{\delta\psi}
\end{equation}
motivates the definition of the conformable derivative: 
\begin{equation}
D_{T}^{(\mu)}\psi(T):=T^{1-\mu}\frac{d\psi}{dT},
\end{equation}
leading to the generalized relaxation law: 
\begin{equation}
D_{T}^{(\mu)}\psi(T)\sim\frac{\delta\mathcal{F}}{\delta\psi}.
\end{equation}

This result gives a first-principles justification for using conformable
calculus in the dynamics of critical systems. The exponent $\mu$
is not a free parameter but encodes the scaling of the relaxation
rate $\Gamma(T)$ due to disorder, statistical structure, and critical
fluctuations.

Therefore, the conformable parameter $\mu$ can serve as a \emph{classification
index} for extended universality classes in nonequilibrium systems.
It provides a bridge between microscopic dynamics and macroscopic
evolution, potentially explaining deviations from classical theory
in real materials.

To consolidate the insights from both classical and conformable frameworks,
we summarize their essential differences in Table \ref{tab:comparison_dynamics}.
This comparison underscores how incorporating fractional noise statistics,
memory effects, and temperature-dependent kinetics extends the classical
relaxation paradigm. The conformable formulation naturally accounts
for anomalous relaxation and long-range temporal correlations, making
it especially well-suited for capturing the complex dynamics observed
near criticality in disordered systems. 

\begin{table}[H]
	\centering
	\caption{Comparison between classical and fractional noise-induced dynamics}
	\label{tab:comparison_dynamics}
	\begin{tabular*}{\linewidth}{@{\extracolsep{\fill}}ll}
		\hline
		\textbf{Classical Dynamics} & \textbf{Fractional / Conformable Dynamics} \\
		\hline
		White noise (uncorrelated) & Fractional noise (long-range correlations) \\
		Local-in-time evolution & Memory-dependent evolution \\
		Standard derivative $\frac{d\psi}{dt}$ & Conformable derivative $T^{1-\mu}\frac{d\psi}{dT}$ \\
		Relaxation time: constant or weakly $T$-dependent & Relaxation time: strongly $T$-dependent via $T^{\mu-1}$ \\
		No history dependence & Dynamics influenced by full past \\
		Markovian dynamics & Non-Markovian dynamics \\
		Debye relaxation & Power-law relaxation, stretched exponentials \\
		Gaussian statistics & Heavy-tailed (Tsallis-like) statistics \\
		\hline
	\end{tabular*}
\end{table}

This framework transforms conformable calculus from a purely mathematical
tool into a physically grounded approach for describing complex systems
with long-range correlations, disorder, and memory. It offers a unified
perspective on fractional kinetics, anomalous relaxation, and thermal
scaling near continuous phase transitions. 

\section*{Acknowledgments:}

The author, José Weberszpil, wishes to express their gratitude to
FAPERJ, APQ1, for the partial financial support.

\section*{Declarations}

\textbf{Funding:} The author received partial financial support from
FAPERJ/APQ1.

\textbf{Conflicts of interest/Competing interests:} The author declares
no conflicts of interest or competing interests related to this work. 

\bibliographystyle{apsrev}
\bibliography{references}

\end{document}